\documentclass{article}
\usepackage{authblk}
\usepackage{multicol}
\usepackage[utf8]{inputenc}
\usepackage{graphicx}
\usepackage{url}
\usepackage{float}
\usepackage{authblk}
\usepackage{placeins}
\usepackage{tabularx}
\usepackage{algorithm2e}
\usepackage{array}

\title{The Contact and Mobility Networks of Mexico City}
\author{
Guillermo de Anda-Jáuregui$^{1,2,3+*}$, Plinio Guzmán$^{4+}$, Oscar Fontanelli$^{5}$, Amilcar Meneses $^{6}$, Alfredo Hernández $^{7}$, Janeth de Anda-Gil$^{5}$, Marisol Flores-Garrido$^{8}$, Maribel Hernández-Rosales$^{5*}$
}

\affil[${1}$]{\footnotesize Computational Genomics Division, Instituto Nacional de Medicina Genómica, Mexico City, Mexico} 
\affil[${2}$]{ Programa de Cátedras CONACYT para jóvenes investigadores, Mexico City, Mexico}
\affil[${3}$]{ Centro de Ciencias de la Complejidad, Universidad Nacional Autónoma de México, México City, México}
\affil[${4}$]{ Independent Science Communication Collective www.estornuda.me}
\affil[${5}$]{ Centro de Investigación y de Estudios Avanzados del Instituto Politécnico Nacional, Unidad Irapuato, Guanajuato, México}
\affil[${6}$]{ Centro de Investigación y de Estudios Avanzados del Instituto Politécnico Nacional, México}
\affil[${7}$]{ Centro de Ciencias Genómicas, Universidad Nacional Autónoma de México, Cuernavaca, México}
\affil[${8}$]{ Escuela Nacional de Estudios Superiores, Universidad Nacional Autónoma de Méxicocampus Morelia, Michoacán, México}
\affil[${+}$]{These authors contributed equally to this work.}
\affil[${*}$]{To whom correspondence must be addressed.}

\date{}

\begin{document}
\maketitle



\paragraph{Subject.}{complex networks, systems biology, data science}


\paragraph{Email.}{gdeanda@inmegen.edu.mx, maribel.hr@cinvestav.mx}
\begin{abstract}
Mexico City, the largest city in Mexico, is also one of the largest cities in the world. It has over 9 million inhabitants and concentrates the vast majority of government and business centers. In this work we describe algorithms that use anonymized location data from mobile devices to construct Mexico City's contact and mobility networks aiming to help the analysis of the city's complexity by understanding movement and physical interaction patterns between its inhabitants. We show the effectiveness and usefulness of our approach by building networks with data collected in February 2020 and performing a general descriptive analysis on them.  We found that contact networks in Mexico City are very sparse, characterized by a largest connected component, and with a heavy-tailed degree distribution. On the other hand, we observed that paths conformed by the highest-degree nodes of mobility networks resemble Mexico City's street network; moreover, we found interesting qualitative differences in the degree distribution of these networks between weekends and weekdays. We present these results along with the release of contact and mobility networks. 
\end{abstract}

Networks have been called a Rosetta stone for complex systems \cite{Farmer1990}. They can represent linked variables and fluid connections and allow for qualitative and quantitative comparisons, providing insights into similarities and differences at different abstraction levels. The power of networks becomes evident when modeling urban dynamics, as cities represent a space that enables all sorts of connections, from people to vehicles, products and ideas. Cities can be seen as a place of networks convergence, a network of networks \cite{Pflieger2010}.\\

In this work we propose an approach to build two different types of networks that model dynamics in Mexico City - a capital city with over 9 million inhabitants and a floating population of over 22 million composed of daily commuters and international visitors. \\

Being the seat of the federal government and concentrating a large fraction of Mexican corporation headquarters, Mexico City is the largest in the country. As it has been proposed for megacities, Mexico City behaves as a complex system: adaptive, evolving, and composed of several layers of interactions that range from infrastructure and transport networks to physical interactions between individuals \cite{Barthelemy2019, Gallotti2019}.\\

We use mobile data from Mexico City to construct contact and mobility networks. In the first case, we create individual-level contact networks using proximity information inferred by the geolocalization data of users in the city.  Contact networks are important because physical contacts of people within the city reflect its organization: its history of urbanization, the layout of its transport infrastructure, as well as the evolution and distribution of its residential, commercial, and industrial zones. The network of physical contacts is perhaps the most fundamental, as it is where human interactions, such as communication (verbal and non-verbal), exchange of goods and services and physical conflict exist. The emergence of higher-order structures and dynamics, such as collective opinion formation and the transmission of infectious diseases, happens on contact networks \cite{Barrat2008}. \\

In the second case, we use both geodata and timestamps to create mobility networks that describe flows between different regions of the city. These networks have weighted edges representing the observed strength in the relationship between regions and capture the specific ranges of urban traveling that citizens create along with their daily routines. Mobility networks have been used to understand the interaction between different city areas and represent a first step towards other computational studies. Centrality on nodes and edges could help to identify both important regions and critical roads in transportation \cite{Huang2015}, community detection could reveal underlying patterns of urban development and growth \cite{Yildirimoglu2018,Liu2015}, and crowded points can help to trace the mobility patterns around points of interest \cite{Du2017,Zhong2015}. Both types of networks could help understand behavioral patterns, associations, preferences and hotspots within the city. \\

Extracting information from mobile data encompasses all the challenges associated with analyzing large datasets, from data storage to the need for highly efficient methodologies that allow for data conversion into different representations - as is our case. In this work, we describe our methodology and build networks using data from February 2020. Then, aiming at showing the efficiency of our approach and the usefulness of the created networks, we perform some general descriptive analysis on them regarding common descriptors, degree distribution and connected components.  Finally, it is worth mentioning that we have included relevant privacy requirements in our design, reaching a balance between model creation and adherence to ethical guidelines. 

\section*{Results}

\begin{figure*}[htp]
    \centering
    \includegraphics[width=.75\linewidth]{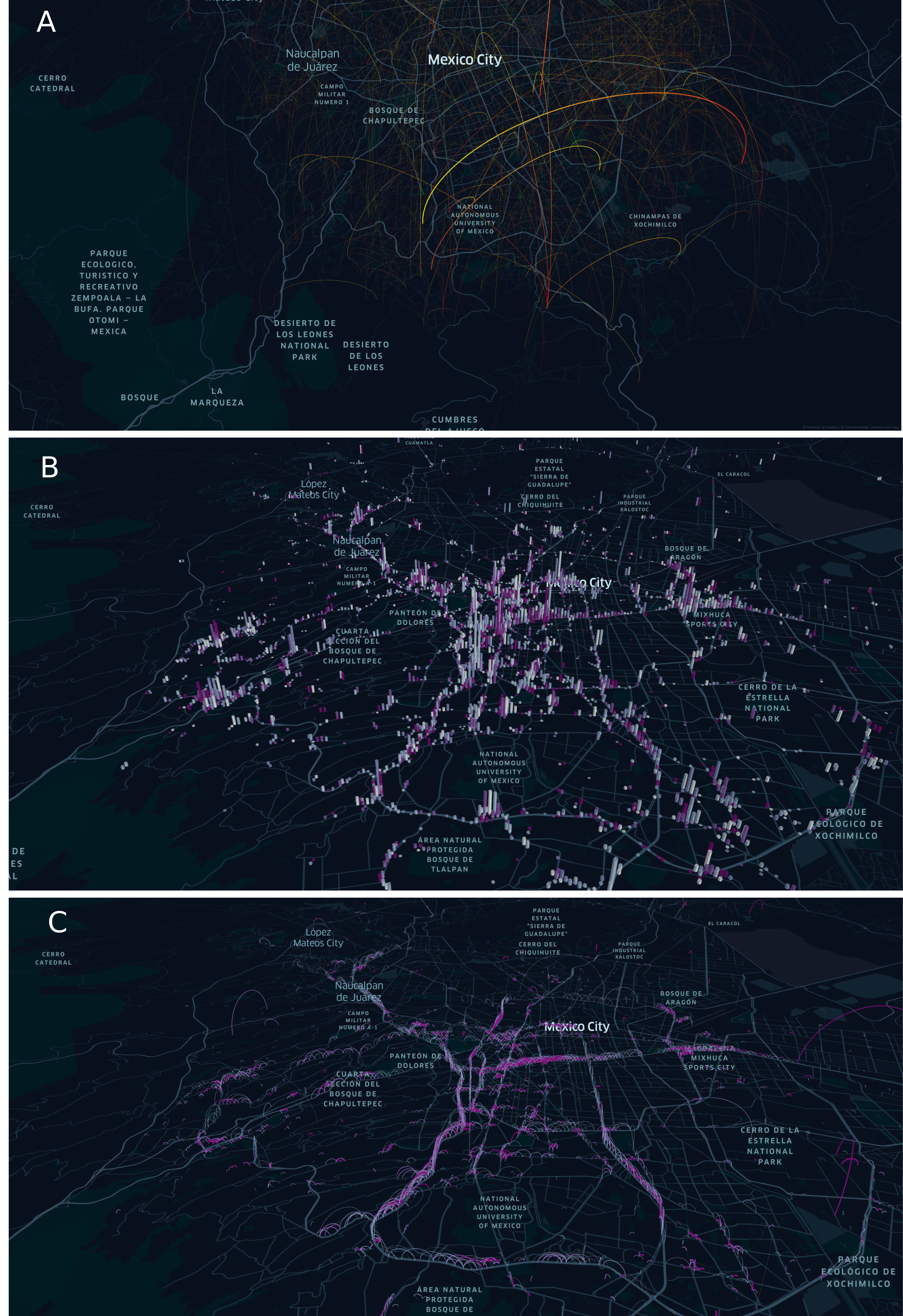}
    \caption{\textbf{Network visualizations}. A) Contact network. Nodes are placed on their nearest origin neighborhood (aggregated to be visible at this scale). Links represent co-localization events throughout the day; thickness represents frequency of contacts between nodes placed on said inferred neighborhood. An arbitrary subset of nodes were selected for this visualization, for illustration purposes.  B) Mobility network. High degree nodes of the mobility network according their strength. C) Mobility network. Edges connecting high-degree nodes.}
    \label{fig:F_03}
\end{figure*}
\begin{figure*}[htb]
    \centering
    \includegraphics[width=0.95\linewidth]{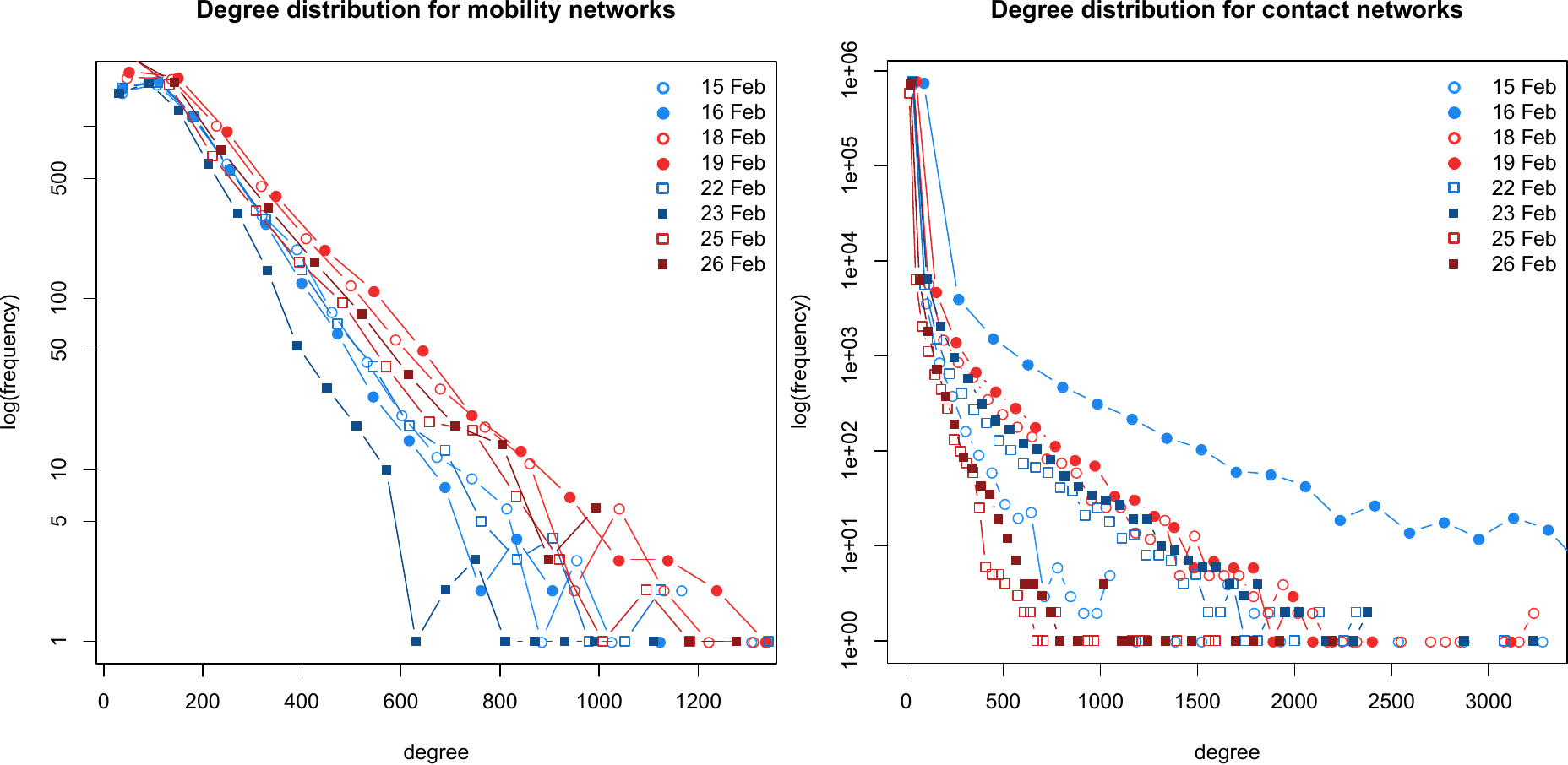}
    \caption{\textbf{Histograms of degree distributions for mobility and contact networks}. Blue lines correspond to weekends and red lines to weekdays. Both histograms are shown in semi-log scale.}
    \label{fig:F_05}
\end{figure*}
\begin{figure*}[htp]
   \centering
    \includegraphics[width=\linewidth]{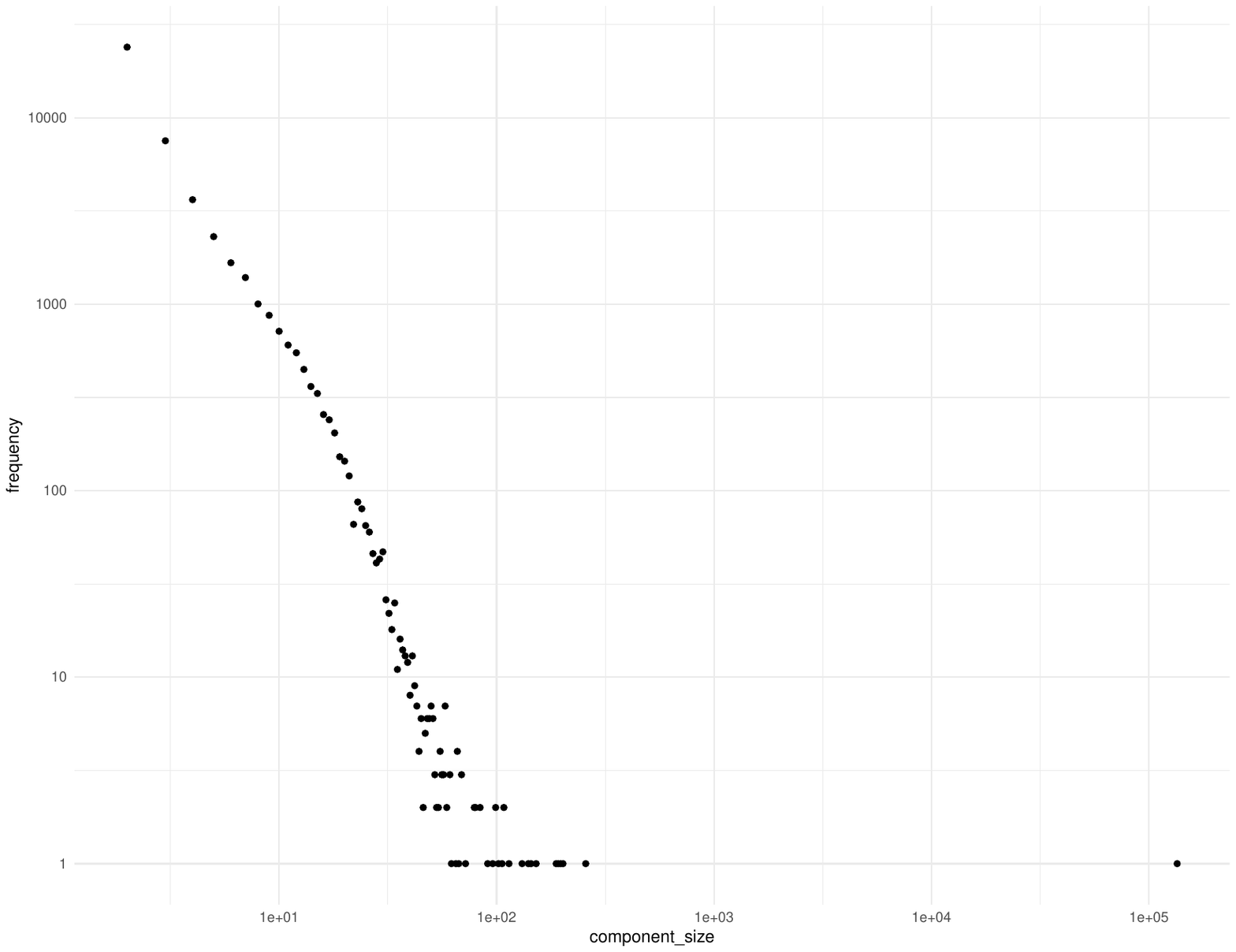}
    \caption{\textbf{Component size distributions of the 2020-02-18 contact network}.  Size distribution for the network's connected components in log scale. The large gap between the largest connected component and the second largest one is evident. Single nodes were removed from the graph.}
    \label{fig:F_04}
\end{figure*}

We have generated a set of contact and mobility networks for Mexico City in February 2020. These networks have been released at doi.org/10.17605/OSF.IO/B6G92.  Basic network descriptors of the networks are shown in Table \ref{T1}. We also calculated the connected components of each network and reported the number of nodes of the three largest ones, as well as isolated nodes (singletons) in Table \ref{T2}. Network visualizations are shown in Figure \ref{fig:F_03}. Additionally, Figure \ref{fig:F_05} shows degree distributions for these networks. \ref{fig:F_04} shows the the long-tailed behavior component size of found in contact networks; as opposed to mobility networks, which are composed of less than 20 connected components.\\

\begin{table*}[htp]
\centering
\caption{Basic descriptors for Mexico City's contact networks. Dates were chosen to have four weekdays (wd) and four weekends (we).}
\label{T1}
\begin{tabular}{llrrrrrr}
  \hline
&\textbf{Date} & {\bf Nodes }& {\bf Edges} &{\bf  Density }&{\bf  Average }&{\bf  Clustering }&{\bf  Average $k$} \\ 
&&&&&{\bf path length}&{\bf coefficient}&\\
  \hline\hline
\textbf{Contact} & 2020-02-15 (we) & 744233 & 1180196 & 0.000004 & 13.50 & 0.30 & 3.17 \\ 
\textbf{Networks} & 2020-02-16 (we) & 767296 & 1180196 & 0.000011 & 13.20 & 0.27 & 8.45 \\ 
 & 2020-02-18 (wd) & 783509 & 2005881 & 0.000007 & 11.46 & 0.28 & 5.12 \\ 
 & 2020-02-19 (wd) & 801419 & 2176754 & 0.000007 & 11.28 & 0.29 & 5.43 \\ 
 & 2020-02-22 (we) & 785043 & 1688023 & 0.000005 & 11.16 & 0.31 & 4.30 \\ 
 & 2020-02-23 (we) & 777879 & 2014671 & 0.000007 & 11.53 & 0.33 & 5.17 \\ 
 & 2020-02-25 (wd) & 590668 & 806429 & 0.000005 & 10.20 & 0.49 & 2.73 \\ 
 & 2020-02-26 (wd) & 727545 & 1169972 & 0.000004 & 11.60 & 0.49 & 3.21 \\
\hline
{\bf Mobility}&2020-02-15 (we) & 5822 & 436901 & 0.013 & 2.60 & 0.20 & 75.0 \\
{\bf Networks}&2020-02-16 (we) & 5813 & 411976 & 0.012 & 2.71 & 0.19 & 70.9 \\
&2020-02-18 (wd) & 5821 & 479283 & 0.014 & 2.64 & 0.20 & 82.3 \\
&2020-02-19 (wd) & 5817 & 490037 & 0.014 & 2.62 & 0.20 & 84.2 \\
&2020-02-22 (we) & 5812 & 423232 & 0.013 & 2.69 & 0.19 & 72.8 \\
&2020-02-23 (we) & 5810 & 354922 & 0.011 & 2.76 & 0.17 & 61.1 \\
&2020-02-25 (wd) & 5813 & 380175 & 0.011 & 2.85 & 0.22 & 65.4 \\
&2020-02-26 (wd) & 5810 & 411596 & 0.012 & 2.77 & 0.18 & 70.8 \\
\hline
\end{tabular}
\end{table*}

\begin{table*}[htp]
\centering
\caption{Connected components in Mexico City's contact and mobility networks}
\label{T2}
\begin{tabular}{lrrrrrr}\hline
&\multicolumn{3}{c}{{\bf Contact Networks}}&\multicolumn{3}{c}{{\bf Mobility Networks}}\\
{\bf Date} & Strong & Rel. size & Singletons &Strong & Rel. size & Singletons \\ 
& Components & largest SC & & Components & largest SC & \\
\hline\hline
2020-02-15 (we) & 49904 & 0.12 & 407625 & 13  & 0.998  & 0\\
2020-02-16 (we) & 50191 & 0.14 & 417502 & 10  & 0.998  & 0\\
2020-02-18 (wd) & 47262 & 0.17 & 432386 & 16  & 0.997  & 0\\
2020-02-19 (wd) & 46617 & 0.19 & 439118 & 10  & 0.998  & 0 \\
2020-02-22 (we) & 49227 & 0.15 & 438402 & 13  & 0.998  & 0 \\
2020-02-23 (we) & 43393 & 0.14 & 447046 & 15  & 0.998  & 0 \\
2020-02-25 (wd) & 33999 & 0.11 & 387236 & 13  & 0.998  & 0 \\
2020-02-26 (wd) & 43588 & 0.17 & 397744 & 13  & 0.998  & 0 \\
\hline
\end{tabular}
\end{table*}

\FloatBarrier


\FloatBarrier

\section*{Discussion}



The two types of networks are capturing two main, complementary components of human behavior in Mexico City. On one hand, we have the \textit{mobility}; that is, the flow of people within the city. On the other hand, we have the \textit{contact} structure, the way in which individual people share the space with each other. By analyzing these networks, we identify some features of this system.\\

In terms of general network structure, we have observed that contact network properties fluctuate for different days, however we have found that they do not exhibit significant changes. For instance, we compared each of these parameters between \textit{weekends} (we) and \textit{weekdays} (wd) networks using a t-test and none of them exhibited significant differences ($p>0.05$).\\

Contact networks exhibit a particular structure in terms of their connected components. Table  \ref{T2} shows descriptors related to these structures, while Figure 3 shows its size distribution.  Meanwhile, mobility networks are much less desegregated in this regard. Analyzing the component distribution of this network, we found that the largest connected component contained 134763 nodes; only  $17.20\%$ of the network's total number of nodes. The second-largest component, in turn, is made of only 257 nodes. Meanwhile, 23907 components are composed of just two nodes. \\

Fig. \ref{fig:F_05} shows histograms in semi-log scale for the degree distributions of the mobility and contact networks. Blue lines correspond to weekends and red lines to weekdays. For contact networks, visual inspection shows that the network is heavy-tailed. This property was further inspected with an algorithm to evaluate power-law distributions in empirical data \cite{Clauset2009}, which indicated that although heavy-tailed, the distribution does not perfectly fit a power-law ($p=0.04$).  Meanwhile, for a specific region of the node degree, histograms for weekends and weekdays for mobility networks fall into two separated groups, indicating two distinct kinds of behaviors. We fitted the linear part of each histogram to a linear function and compared the two groups of estimated slopes (weekends vs. weekdays) via a $t$-test. We got a $p-value$ very close to zero, confirming the observations that these two groups of distributions are significantly different. We do not observe this separation in the distributions for contact networks. Here, consecutive days tend to have similar distributions but without a weekend-weekday separation. Observe that there is a particular day (February 16th) where the degree distribution of the contact network has a much heavier tail than the rest. \\

To study the regions with highest number of visits during the days of study, we analyzed the degree of the nodes. For illustration purposes, we show in Fig. \ref{fig:F_03} B and C visualizations of the mobility network for February 18. First, we show on panel B the highest-degree nodes of this network over a map of Mexico City, where each bar's color and height represent each node's strength (sum of weights of incoming or outgoing edges). On panel C, we show the edges connecting the highest-degree nodes;  it is worth highlighting that these edges resemble the city's street network. \\



In this work, we show a simple way in which device position data can be leveraged to construct contact and mobility networks for Mexico City. Basic network descriptors and visualizations show characteristic topologies for these networks. For contact networks, we see that this topology, although fluctuating, remains relatively stable through the different dates analyzed. Based on its construction, we believe that these contact networks adequately capture the heterogeneous nature of physical contacts in Mexico City. However, there seems to be a clear separation for mobility networks between weekends and weekdays, suggesting qualitative differences in mobility patterns depending on the day of the week.  \\

We focused on reconstructing a contact network that captures the contacts observed throughout a single day as a first approach. The strategy that we propose can be adapted to analyze narrower time windows. That scenario could be of interest to identify changes in connectivity patterns that occur during a given day; for instance, contacts observed during business hours or late at night could change. It has been shown that different levels of time aggregation may obscure or reveal different patterns within a network \cite{Sapiezynski2019}. Correspondingly, larger time windows for the construction of mobility networks could give valuable and interesting insights into large-scale mobility patterns for people in the city. \\

Due to the computational cost involved in all steps of this analysis pipeline, it has been unfeasible to process a more significant number of dates. However, we believe that it is important to identify whether the topological structure of contact networks remains throughout a larger period (for instance, a year). Similarly, it will be important to verify the difference between weekday and weekend degree distribution for mobility networks with a larger number of observations. \\

While the overall structure of contact networks is retained on different days, an open question is whether well-documented changes in mobility patterns lead to what would be, in essence, a rewiring of the network. For instance, weekday commuters could interact with a set of people at work and have the same number of contacts with different people on the weekend. A particular question that we plan to tackle is whether pandemic mobility restrictions have successfully reduced the number of contacts within the city or have only led to a somewhat degree-preserving rewiring of the network. Similarly, it will be interesting to see whether the properties of mobility networks change after these restrictions, indicating modifications in the mobility patterns of people in the city due to the pandemic. \\

We must acknowledge that the particular motivation of this work is in the context of the current COVID-19 pandemic. While we believe that there are many applications for this network analysis, we are particularly interested in its use for the development of epidemiological and other public policy decision-making models. With this in mind, we are providing these networks as an open dataset for any researcher or policymaker interested in their use.\\

The same strategy used for the reconstruction of these networks can be applied to other cities in Mexico and Latin America. Since mobile devices are widely used throughout urban areas, the data is available. It would be interesting to see if the structures observed for Mexico City are preserved in other cities throughout the continent.  It is our group's intention to continue increasing the number of available contact and mobility network data in this open resource. Our goal is to be able to provide these datasets for all major metropolitan areas in Mexico, for longer periods of time. We believe that such open datasets can be useful for decision makers in Mexico and in other countries in the region, in which access to such resources and the associated computational power to manage them may not be easily obtained. Our dataset is hosted in OSF (doi.org/10.17605/OSF.IO/B6G92), through which future updates will be made available. 

\section{Materials and methods}

\subsection{Data acquisition and pre-processing}

Device location data for February 2020 was provided by a company that aggregates anonymized cell phone location data. Each data record comprises a unique anonymized user id, a timestamp, horizontal accuracy, and a latitude and longitude pair. The set of unique user ids is referred to as the panel. 

Efficiently analyzing large datasets of individual locations can be done by bucketing locations \cite{Sahr2003} into area partitions. We first sorted the dataset by administrative region to select a subset of data, then partitioned that subset of data over a hexagonal grid. 
We chose the H3 geospatial indexing system\footnote{\url{https://github.com/uber/h3}}, aligned to the EPSG 4326 Coordinate Reference System (CRS). H3 tiles Earth with non-overlapping hexagons at 16 different linearly hierarchical resolutions. Given a latitude and longitude point, the API returns an index of the containing H3 cell at a particular resolution. This grid system provides uniform spacing between cells and, because all neighbors in a hexagonal grid are equidistant, facilitates analysis of movement across cells. 

For this study, we selected records that appear within a bounding box containing the shapefile of Mexico City published by the Mexican National Institute of Statistics and Geography (INEGI). After exploring the distribution of horizontal accuracy, we filtered out pings with a horizontal accuracy larger than 100 m to reduce the number of false-positive links between devices in the networks. 

\subsection{Contact network reconstruction}

\begin{figure*}[htp]
    \centering
    \includegraphics[width=0.7\linewidth]{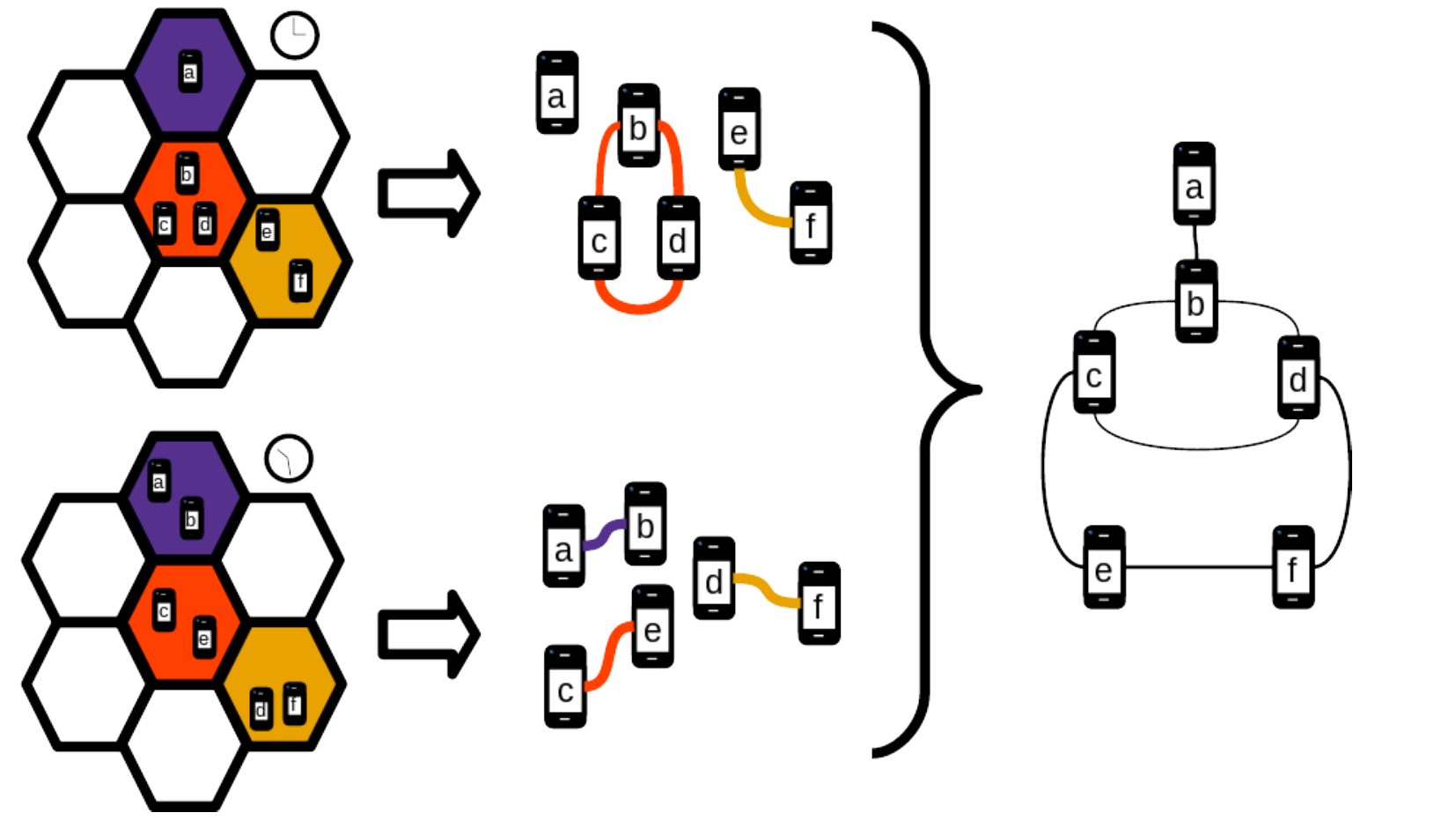}
    \caption{\textbf{Schematic representation of our contact network construction method}. Devices at each time slice t, devices occupying the same position in the H3 grid are connected (formally, a bipartite device - position network is generated and projected to the network; see methods). The networks for all time slices that compose the time period of interest (eg: one day) are then merged to form a single contact network. }
    \label{fig:F_01}
\end{figure*}

We reconstructed a network for each of the analyzed days, using nodes to represent devices (as a proxy for people) and links represent \textit{physical proximity contacts}.  We define the occurrence of a physical proximity contact when devices are simultaneously in the same hexagon of the grid previously described during the day of analysis. 

The algorithm \ref{Alg1} shows the procedure used to construct the contact network. Figure \ref{fig:F_01} shows a schematic representation of the network construction methodology. 

\begin{algorithm}[H]
 \KwData{Device locations.}
 \KwResult{Contact network.}
 \ForEach{$day$}{
  break $day$ into 144 time windows $tw$ of 600 seconds;\\
    \ForEach{$tw$}{
       Make a bipartite network linking devices to all positions they occupy during $tw$. \\
       Project the network to the device layer.\\
       Return $Gd_{tw}$, a network in which devices are linked through their (aggregated) co-localization events.
    }
    Merge all $Gd_{tw}$ to obtain $Gd_{day}$, the contact network for a given day.
 }
 \caption{Algorithm for contact network construction.}
 \label{Alg1}
\end{algorithm}
\smallskip

By aggregating networks at the daily level, we can capture the general connectivity patterns of the city without risking exposing individual mobility patterns. Thus, we would like to state that no individual mobility pattern was explicitly or implicitly traced or needed to construct the networks.

\subsection{Mobility network reconstruction}

\begin{figure*}[htp]
    \centering
    \includegraphics[width=0.5\textwidth]{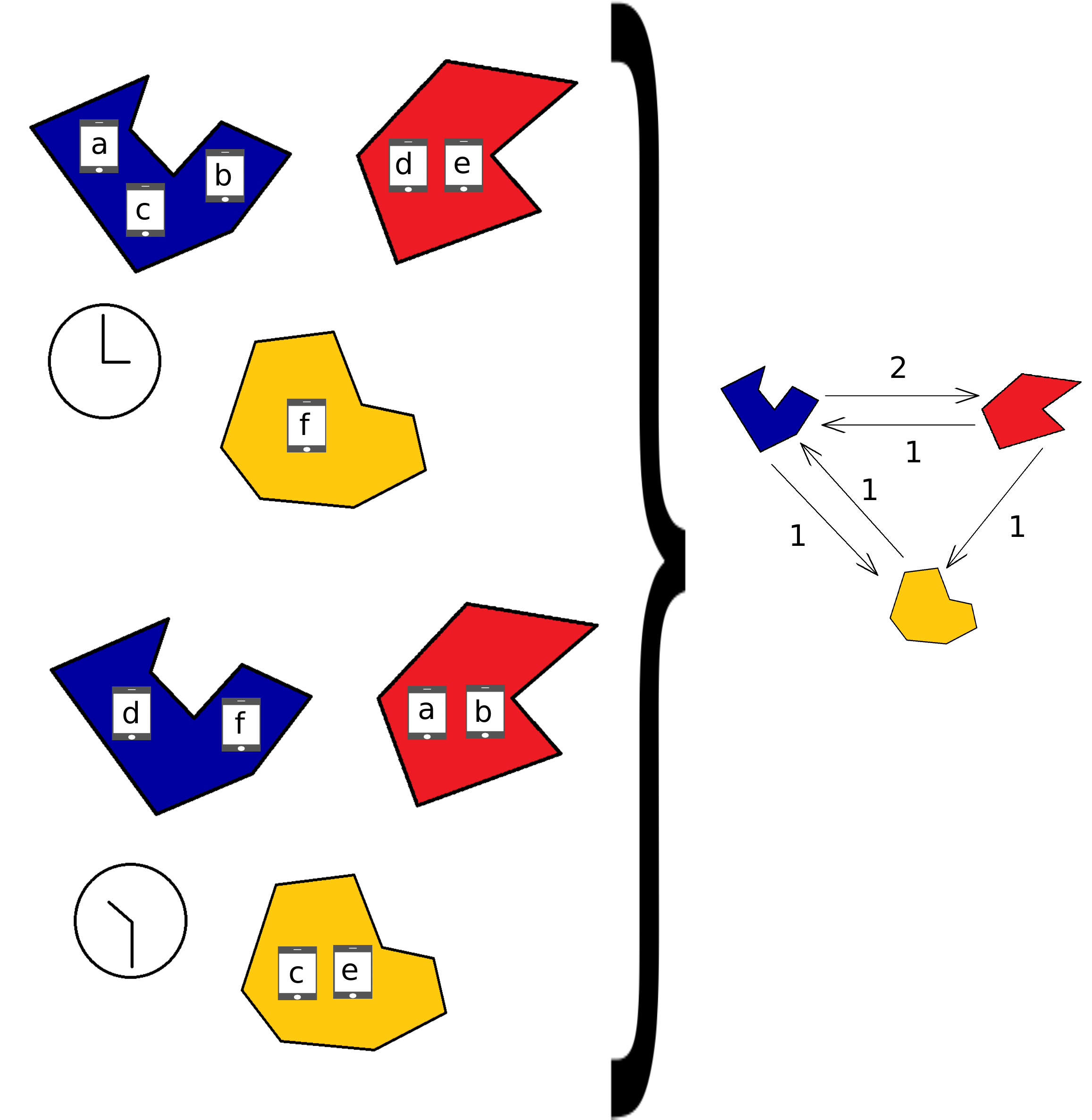}
    \caption{\textbf{Schematic representation of our mobility network construction method}. Polygons are AGEBs which contain certain devices at time 1, then these devices change their locations at time 2. For each ordered pair of AGEBs we count the number the devices that moved from one to the other and we add this quantity to the the weight of the respective edge. }
    \label{fig:F_02}
\end{figure*}

In mobility networks, nodes are geographic areas defined by INEGI called AGEBs (Basic Geo-statistic Area). These units represent the basic territorial subdivision unit, and they cover the totality of the territory. Edges in the mobility network are weighted and directed; the weight of the edges represents the number of devices that moved from one node to the other during a time window of 10 minutes. We construct the mobility network using algorithm \ref{Alg2}, and we show in figure \ref{fig:F_02} a schematic representation of this methodology.

\begin{algorithm}[H]
 \KwData{Device locations, list of AGEBs $\{a_i\}$ with geographic location.}
 \KwResult{Mobility network (weighted and directed). }
 \ForEach{$day$}{
  break $day$ into 144 time windows $tw$ of 600 seconds;\\
  Create a complete directed network where nodes are AGEBS $\{a_i\}$ and all edges have weight $w_{ij}=0$. \\
    \ForEach{$tw$}{
       Count number of devices that where in AGEB $a_i$ at time $tw-1$ and were in AGEB $a_j$ at time $tw$. Call this number $k_{ij}(tw)$.\\
       Add this number to the weight of the edge from $a_i$ to $a_j$,\\
       $$
       w_{ij} = w_{ij}+k_{ij}(tw).
       $$
    }
    Delete all edges with weight equal to zero.
 }
 \caption{Algorithm for mobility network construction.}
 \label{Alg2}
\end{algorithm}



\enlargethispage{20pt}

\paragraph{Ethics.}{Networks has been processed to substitute the original anonymous ids with new numeric ids to guarantee that nodes cannot be reidentified.}

\paragraph{Data availability.}{Networks described in this manuscript are available at \textit{graphml} file, available at: doi.org/10.17605/OSF.IO/B6G92 (contact networks) and doi.org/10.17605/OSF.IO/86F3C (mobility networks) .}

\paragraph{Funding}{This project was supported by the Fondo Conjunto de Cooperaci\'on M\'exico-Uruguay from the Agencia Uruguaya de Cooperaci\'on Internacinal and the Agencia Mexicana de Cooperaci\'on Internacional para el Desarrollo.}

\paragraph{Acknowledgements}{We would like to thank Jens Steuck for technical support, as well as the Bioinformatics Group of the University of Leipzig for providing computational infrastructure during the development of this project.}


\bibliographystyle{plain}
\bibliography{refs}

\end{document}